\documentclass[usenatbib,usegraphicx]{mn2e}

\def\md{--~~\,}

\def\lesssim{\mathrel{\hbox{\rlap{\hbox{\lower4pt\hbox{$\sim$}}}\hbox{$<$}}}}

\title[Postmortem of SN 2001ig]{A postmortem investigation of
the Type~IIb supernova~2001ig}
\author[S. D. Ryder, C. E. Murrowood and R. A. Stathakis]{Stuart D.
Ryder$^{1}$\thanks{E-mail: sdr@aao.gov.au}, Clair E. Murrowood$^{1,2}$
and Raylee A. Stathakis$^{1}$\\
$^{1}$Anglo-Australian Observatory, P.O. Box 296, Epping, NSW 1710, Australia\\
$^{2}$School of Mathematics \& Physics, Private Bag 21, University of Tasmania,
Hobart, TAS 7001, Australia}
\begin{document}

\date{Accepted 2006 March 13. Received 2006 March 10; in original form
2006 February 27}

\pagerange{\pageref{firstpage}--\pageref{lastpage}} \pubyear{2006}

\maketitle

\label{firstpage}

\begin{abstract}

We present images taken with the GMOS instrument on Gemini-South, in
excellent ($<$0.5~arcsec) seeing, of SN~2001ig in NGC 7424,
$\sim$1000~days after explosion. A point source seen at the site of
the SN is shown to have colours inconsistent with being an H\,{\sc
ii}~region or a SN~1993J-like remnant, but can be matched to a late-B
through late-F supergiant with $A_{V}<1$. We believe this object is the
massive binary companion responsible for periodic modulation in mass
loss material around the Wolf-Rayet progenitor which gave rise to
significant structure in the SN radio light curve.

\end{abstract}

\begin{keywords}
stars: evolution -- supernovae: general -- supernovae: individual: SN 2001ig
-- binaries -- galaxies: individual: NGC 7424.
\end{keywords}

\section{Introduction}
\label{s:intro}

Supernovae (SNe) of Type II, Ib, and Ic are all now recognised as
arising from the core collapse of a massive ($>8$~M$_{\odot}$) star, but
with progressively less H and He in their outer layers. As a result,
early spectra from Types Ib and Ic lack any H features, while Type Ic
even lack any He features \citep{avf97}. The rare class of Type IIb
supernovae undergo a spectral transition from Type II to Type Ib as
their hydrogen recombination lines fade.  As such, they could offer
important clues about the nature of this mass loss, via stellar winds
or mass transfer in binary systems.

Two of the nearest examples of Type IIb SNe have also been found to be
luminous at radio and X-ray wavelengths, indicative of interaction with
a dense, pre-existing circumstellar medium (CSM): SN 1993J in M81
\citep{93jrad,93jx}, and SN~2001ig in NGC~7424 (\citealt{sdr04},
hereafter Paper~I; \citealt{sr02}).
The multi-frequency radio light curves of SN~2001ig showed clear
modulations with a period $\sim150$~days, due to corresponding changes
in the CSM density. While thermal pulsations in the core of a single
AGB star progenitor could in principle generate mass-loss shells with
the appropriate spacing, Paper~I argued instead that a massive
binary companion in an eccentric orbit about a Wolf-Rayet (WR)
progenitor would yield the necessary mass-loss enhancements near
periastron. The `pinwheel' dust nebulae formed from the colliding stellar
winds have even been imaged directly in some Galactic WR binary systems
\citep{tut03}.

The idea that mass-transfer in binary systems could induce the
stripping necessary to account for the Type IIb phenomenon received a
boost with the unmasking of a hot massive binary companion to
SN~1993J, initially via photometry of pre-supernova imaging
\citep{ahr94} and ultimately via its signature in late-time
ultraviolet spectra \citep{nat04}. We predicted (Paper~I) that if the
progenitor of SN~2001ig had a binary companion, then it too should
become apparent, once the optical remnant had faded. In this Letter we
report the results of multi-colour ground-based optical imaging, under
excellent seeing conditions, which appear to show just such an object.

\section[]{Observations and Data Reduction}
\label{s:data}

Imaging of NGC~7424, which takes in the location of SN~2001ig, was
conducted in queue mode with the Gemini Multi-Object Spectrograph
(GMOS; \citealt{gmos04}) attached to the Gemini-South Telescope for
programme GS-2004B-Q-6 (PI: S. Ryder). Each GMOS image yields
an unvignetted field of view $\sim$5.5~arcmin on a side, at
0.073~arcsec~pix$^{-1}$.  Observing was carried out on the night of
2004 Sep 13 UT when the seeing was 0.6--0.8~arcsec and conditions were
photometric, and repeated on 2004 Sep 14 UT when the seeing was
0.35--0.45~arcsec but cirrus and thin cloud were present.  On the
second night a series of $5 \times 540$~s exposures in the Sloan
Digital Sky Survey (SDSS) $u^{\prime}$ filter, $3 \times 240$~s in
$g^{\prime}$, and $4 \times 530$~s in $r^{\prime}$ were obtained, with
each exposure offset by at least 5~arcsec spatially to allow
filling-in of the inter-CCD gaps in GMOS. Images of the photometric
standard field T~Phe \citep{lan92} were obtained as part of the
baseline calibration for the first night.

The data were reduced and combined using V1.8.1 of the {\sc gemini}
package within {\sc iraf}\footnote{{\sc iraf} is distributed by the
National Optical Astronomy Observatories, which are operated by the
Association of Universities for Research in Astronomy, Inc., under
cooperative agreement with the National Science Foundation.}. A master
bias frame (constructed by averaging with 3-$\sigma$ clipping a series
of bias frames) was subtracted from all raw images in lieu of overscan
fitting and subtraction. Images of the twilight sky in each filter
were used to flatfield the images and suppress a prominent `brick
wall' pattern, particularly in $u^{\prime}$. The dithered galaxy
images in each filter were then registered and averaged together with
the {\sc imcoadd} task to eliminate the inter-CCD gaps, bad pixels,
and cosmic rays.


In order to precisely locate SN~2001ig in our GMOS images,
observations of SN~2001ig taken 6~months after outburst were extracted
from the ESO Science Archive. SN~2001ig was imaged with the FOcal
Reducer/low dispersion Spectrograph (FORS2) on the Very Large
Telescope UT4 during the night of 2002 June~16 UT as part of programme
69.D-0453(B) (PI: E.~Cappellaro). Exposures of 90~s in $B$, 60~s in
$V$, and 15~s in $R$ with a pixel scale of 0.252~arcsec~pix$^{-1}$
were processed using {\sc iraf} in a similar manner to the GMOS
data\footnote{A full-colour representation of these data can be seen
at http://www.eso.org/outreach/press-rel/pr-2004/phot-33-04.html.}.

\begin{figure*}
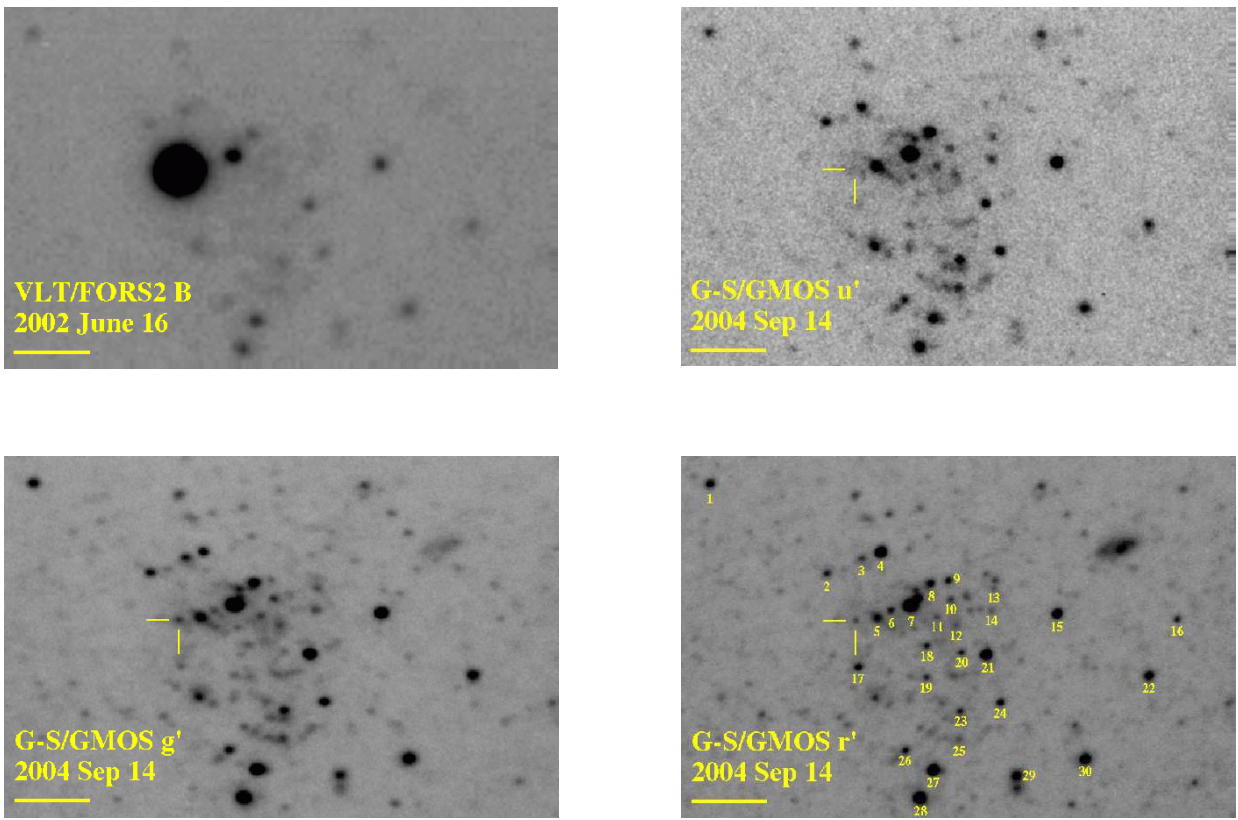

\vspace{12cm}
\includegraphics{vlt_B.cps}
\includegraphics{gmos_u.cps}
\includegraphics{gmos_g.cps}
\includegraphics{gmos_r.cps}
  \caption{Images of the region immediately surrounding
   SN~2001ig in NGC~7424. North is to the top, with East to the left, and
   the line in the lower-left corner of each image is 5~arcsec in length.
   The two perpendicular lines in each of the GMOS images correspond to
   the centroid of the SN as measured from the FORS2 image.
   The numbers immediately below (or to the right of) stars in the
   $r^{\prime}$ image identify the field stars for which colours have been
   measured in Table~\protect{\ref{t:phot}}, and plotted in
   Fig.~\protect{\ref{f:colours}.}}
\label{f:images}
\end{figure*}

A section covering some 36~arcsec $\times$ 23~arcsec surrounding
SN~2001ig was extracted from the final GMOS images in each filter, and
a matching section extracted from the FORS2 images. These image
sections were then all aligned to a common coordinate system and image
scale using the {\sc geomap} and {\sc geotran} tasks within {\sc iraf}
by fitting to 15 stars in common. Figure~\ref{f:images} shows the
FORS2 $B$, and GMOS $u^{\prime}$, $g^{\prime}$, $r^{\prime}$ images of
the neighbourhood of SN~2001ig. Though the SN appears saturated in all
the FORS2 images, profile-fitting yielded a centroid position
consistent to 0.3~resampled pixels which, including the r.m.s. of the
{\sc geomap} fitting, results in an overall positional uncertainty of
0.03~arcsec in each axis. As Fig.~\ref{f:images} indicates, the
position of the SN on the FORS2 image is coincident with a faint,
point-like source in both the $g^{\prime}$ and $r^{\prime}$ images,
but there is no counterpart in the $u^{\prime}$ image.  Rings and arcs
of diffuse nebulosity are much more apparent in the $u^{\prime}$
image, and the SN position falls on the northern rim of one such
arc. If there is an ultraviolet counterpart to the source seen in the
$g^{\prime}$ and $r^{\prime}$ bands, then it is unfortunately hidden
within this emission.

\section[]{Photometry}
\label{s:phot}

Aperture photometry of the T~Phe observations on 2004~Sep~13 was
carried out using the {\sc phot} task within {\sc iraf}, an aperture
radius of 2.6~arcsec (35~pix), and sky level from the mode of pixels
at radii between 2.9--3.7~arcsec (40--50~pix). While only stars A, C,
and D from \citet{lan92} were within the GMOS field, \citet{jas05}
have determined magnitudes in the SDSS system for these and three
other stars in the same field, which have enabled us to derive
independently the extinction, zero-points, and colour terms in
$u^{\prime}$, $g^{\prime}$, and $r^{\prime}$. We note this is the
first time such quantities have been determined for GMOS on
Gemini-South utilising calibrated SDSS photometry, rather than relying
upon theoretical \citep{fuk96} or empirical \citep{jas02,kar05}
transforms from the Johnson colours. For the T~Phe observations, we
find:

\begin{displaymath}
u^{\prime} = (25.14\pm0.04) - 2.5\log(C/t) - 0.34X
              + 0.15(u^{\prime} - g^{\prime})
\end{displaymath}

\begin{displaymath}
g^{\prime} = (28.54\pm0.01) - 2.5\log(C/t) - 0.12X
              + 0.05(g^{\prime} - r^{\prime})
\end{displaymath}

\begin{displaymath}
r^{\prime} = (28.53\pm0.03) - 2.5\log(C/t) - 0.11X
              + 0.05(g^{\prime} - r^{\prime})
\end{displaymath}

\noindent
where $C$ = integrated counts within 2.6~arcsec aperture radius, $t$ is the
exposure time, and $X$ is the airmass.

Using these relations, we have been able to `bootstrap' a calibration
for the Sep~14 data from the Sep~13 data, by determining mean offsets
in each filter between the 2 nights for large-aperture measurements
of 10 isolated field stars in each image. Due to the semi-crowded
nature of the field shown in Fig.~\ref{f:images}, measurements of
these same 10~stars in apertures ranging from 0.4~arcsec out to
2.6~arcsec have also been obtained, allowing an empirical aperture
correction to be applied in each filter when contamination by
neighbouring stars in the large aperture might be a problem.

The magnitudes and colours for 30~stars in the neighbourhood of
SN~2001ig (as identified in Fig.~\ref{f:images}) are presented in
Table~\ref{t:phot}. A few of these stars had no clear counterpart in
the $u^{\prime}$ image for {\sc phot} to centroid on, so a $3\sigma$
upper limit of $u^{\prime}>26.1$ has been adopted following the
approach of \citet{ms05}. The one exception to this is for the site
of SN~2001ig itself, which sits within a region of diffuse nebulosity,
and an aperture-corrected upper limit on the measured flux within a
0.4~arcsec aperture is $u^{\prime}>24.9$.

\begin{table}
\centering
\caption{Photometry for objects in the neighbourhood of
SN~2001ig.\label{t:phot}}
\begin{tabular}{rcrrrr}
\hline
Star & $g^{\prime}$ & ($u^{\prime} - g^{\prime}$) & ($g^{\prime} -
r^{\prime}$) & $A_{V}$ & Sp. Type \\
\hline
 1 &  23.32 &   0.70   &   0.13   &   1.2   &    B6 \\
 2 &  23.62 &   0.04   &   0.03   &   1.7   & $<$O5 \\
 3 &  23.51 &   0.26   & $-0.07$  & $\sim$3 & $<$O5 \\
 4 &  23.51 & $>2.99$  &   1.50   & $<$3    & $>$G2 \\
 5 &  23.12 & $-0.13$  &   0.27   &   0.8   &    B3 \\
 6 &  24.18 &   0.39   &   0.47   & $\sim$3 & $<$O5 \\
 7 &  21.54 &   0.08   &   0.14   &   1.5   & $<$O5 \\
 8 &  23.03 & $-0.15$  & $-0.12$  &   1.2   & $<$O5 \\
 9 &  24.98 & $>1.29$  &   1.29   & $<$5    & $<$K4 \\
10 &  23.84 &   0.63   & $-0.08$  &   0.4   &    B9 \\
11 &  24.12 &   0.03   &   0.33   & $\sim$3 & $<$O5 \\
12 &  24.22 &   0.37   &   0.33   &   2.5   & $<$O5 \\
13 &  24.39 &   0.09   & $-0.01$  &   1.9   & $<$O5 \\
14 &  24.23 & $-0.23$  & $-0.43$  & $\sim$0 &    B1 \\
15 &  22.34 &   0.13   & $-0.16$  &   0.6   &    B2 \\
16 &  25.95 & $>0.19$  &   1.70   &   \md   &   \md \\    
17 &  24.77 & $>1.54$  &   1.25   & $<$5    & $<$K4 \\
18 &  25.01 &   0.68   &   1.04   & $<$4.5  & $<$K0 \\
19 &  25.16 & $>1.08$  &   0.71   & $<$3    & $>$B1,$<$G3 \\
20 &  25.69 & $>0.48$  &   1.42   & $<$5    & $<$K8 \\
21 &  22.45 &   1.41   &   0.42   &   1.9   &    A0 \\
22 &  23.02 &   0.69   &   0.14   &   1.2   &    B5 \\
23 &  23.49 & $-0.06$  & $-0.37$  & $\sim$0 &    B3 \\
24 &  23.28 &   0.22   & $-0.23$  &   0.2   &    B5 \\
25 &  23.90 & $-0.10$  & $-0.45$  & $\sim$0 &    B2 \\
26 &  23.67 &   0.31   & $-0.03$  &   0.9   &    B3 \\
27 &  22.30 &   1.12   &   0.35   &   1.6   &    B9 \\
28 &  22.19 &   0.82   &   0.15   &   1.0   &    B8 \\
29 &  23.53 &   1.23   &   0.61   &   2.6   &    B6 \\
30 &  22.63 &   0.78   &   0.20   &   1.4   &    B6 \\
SN &  24.14 & $>0.76$  &   0.11   & $<$1.0  & $>$B7,$<$F8 \\
\hline
\end{tabular}
\end{table}

\begin{figure}
\includegraphics[width=84mm]{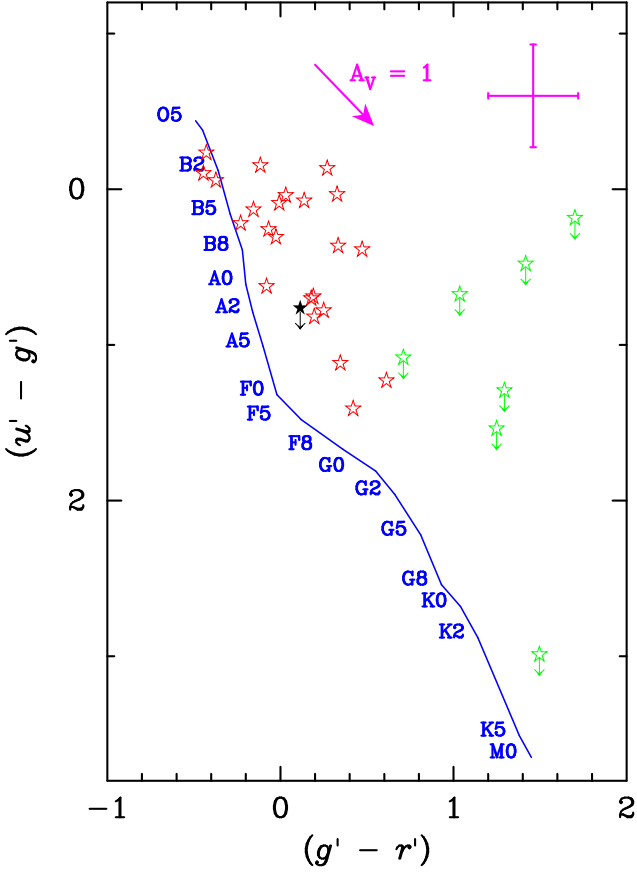}
  \caption{Two-colour diagram in the SDSS system for field stars labeled in
   Fig.~\protect{\ref{f:images}} (open symbols) as well as SN~2001ig itself
   (filled symbol).
   Objects below the detection limit in $u^{\prime}$ are shown with upper
   limits in $(u^{\prime} - g^{\prime})$. The locus of unreddened colours
   for supergiants stars is shown with their spectral type, as well as the
   reddening vector and indicative error bar [$\pm0.33$ in $(u^{\prime} -
   g^{\prime})$, $\pm0.26$ in $(g^{\prime} - r^{\prime})$].}
\label{f:colours}
\end{figure}

The colours of these 30~objects and the counterpart to SN~2001ig are
plotted in Fig.~\ref{f:colours}. Also plotted in this diagram is the
locus of unreddened colours for supergiant stars, transformed from
the Johnson colours \citep{aq4,lang91} to the SDSS system \citep{jas02};
the reddening vector equivalent to 1~mag of extinction in the $V$-band
\citep{fan99}; and an indicative error bar on each point which combines
the uncertainties in the Sep~13 calibration, the bootstrapping to the
Sep~14 data, the {\sc phot} measurement, and any aperture correction.

As Fig.~\ref{f:colours} indicates, stars in the neighbourhood of
SN~2001ig are all consistent (to within the errors or colour limits)
with being supergiants, or possibly young stellar clusters, suffering
varying amounts of reddening. The reddest objects in $(g^{\prime} -
r^{\prime})$ are so red that they are not detectable in $u^{\prime}$.
The extremely red colours of 4, 9, 16, and the background disk galaxy
in Fig.~\ref{f:images} suggest the presence of a foreground dust lane
running from just north, to the west of SN~2001ig. Star~17, just
3~arcsec south of SN~2001ig, is also absent in $u^{\prime}$ but is
intrinsically much redder than the SN. Where possible, we have
`de-reddened' each object back on to the supergiant locus in
Fig.~\ref{f:colours}, and tabulated the estimated extinction and
inferred spectral type in Table~\ref{t:phot}.

\section[]{Discussion and Conclusions}
\label{s:disc}

\subsection{Stellar companion or nebula?}

De-reddening from the blue limit on $(u^{\prime} - g^{\prime})$ for
SN~2001ig indicates an intrinsic spectral type for the companion of B7
or later. However, the $(g^{\prime} - r^{\prime})$ colour and its
uncertainty requires that it be no later than F8 with no reddening. In
addition to the colours, we can use the absolute luminosity to attempt
to constrain the spectral type, though this is subject to the usual
distance uncertainties. For a distance of 11.5~Mpc \citep{nbg88} and
$A_{V}<1$ then $M_{u^{\prime}}>-7.0$, which requires a spectral type
of B0 or later according to the stellar luminosity data and transforms
used in Sect.~\ref{s:phot}. Similarly, $-6.2 > M_{g^{\prime}} > -7.4$
constrains the type to be F8 or earlier, while $-6.3 > M_{r^{\prime}}
> -7.1$ implies a type between A1 and K8. Thus, both the colour and
luminosity constraints are consistent in arguing for the presence of
a late-B through late-F supergiant at the location of SN 2001ig.

While only the $u^{\prime}$ image shows much nebulosity in this
region, the effect of a foreground or background H\,{\sc ii}~region
needs to be considered. Aperture photometry of selected locations
around a prominent `bubble' nebula 40~arcsec east of the SN has been
performed to determine the SDSS colour characteristics of a
low-metallicity H\,{\sc ii}~region. While their $(g^{\prime} -
r^{\prime})$ colours are broadly similar to the SN, all were found to
have $(u^{\prime} - g^{\prime}) \lesssim 0$, much bluer than the SN or
almost every other object in its vicinity. Those locations which
included an ionising star/cluster have colours consistent with $A_V =
1-2$~mag O~stars, while those which are purely nebulous have
$(u^{\prime} - g^{\prime}) \sim -1$, so are too blue to be stars.

\begin{figure}
\includegraphics[height=84mm,angle=270]{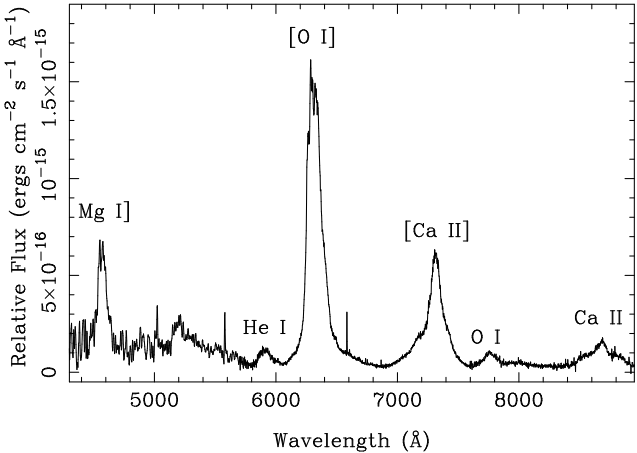}
  \caption{Composite spectrum of SN~2001ig obtained 285~days after
  explosion with the RGO Spectrograph on the AAT, combining a 3600~s
  exposure in the red and 600~s in the blue. The major emission lines
  are identified following \citet{mat00}.}
\label{f:rgospec}
\end{figure}

Another potential contaminant could be the supernova remnant (SNR)
itself.  A decade after SN~1993J was discovered, its spectrum was
still dominated by broad, box-shaped emission features
\citep{nat04}. No spectra of SN~2001ig have yet been
published, so we present in Fig.~\ref{f:rgospec} a spectrum obtained
on 2002~Sep~14 UT with the Anglo-Australian Telescope.  A detailed
spectral analysis is beyond the scope of this Letter, but we note that
this spectrum bears strong similarities to spectra of SN~1993J at a
similar epoch (day~286 in \citealt{mat00}), with the notable absence
of a broad H$\alpha$ component.

In order to assess the colour evolution of a Type~IIb SNR in the SDSS
system, we have performed synthetic photometry on the late-time
spectra of \citet{mat00}, using the response curves of \citet{jas02}.
The zero-points were calibrated by matching the results on a flux-calibrated
spectrum of the fundamental SDSS standard ${\rm BD}~+~17^{\circ}4708$
\citep{bg04} to its defined values. The $(g^{\prime} - r^{\prime})$ colour
of SN~1993J evolved from 1.0 at day~881, to 0.7 at day~976, and 0.6 by
day~1766. It was not until day~2000 that it reached a colour similar
to that observed in SN~2001ig, though dereddening by an amount
corresponding to $A_V = 0.6$~mag \citep{mat00} would bring it just
within range of the colour uncertainty on SN~2001ig by
day~1766. Extended ultraviolet spectral coverage after day~2000 allows
us to constrain $(u^{\prime} - g^{\prime}) \lesssim 0.6$ at this epoch
for SN~1993J, which is even bluer than the blue limit on
SN~2001ig. Thus, even if SN~2001ig had evolved significantly faster
than SN~1993J (though Fig.~\ref{f:rgospec} gives us no reason to
expect this), its SNR might never become blue enough in $(g^{\prime} -
r^{\prime})$ [or red enough in $(u^{\prime} - g^{\prime})$] to fully
account for the properties of the object we see at the location of
SN~2001ig.

\subsection{The progenitor systems of Type IIb supernovae}

SN~2001ig is thus perhaps the second Type IIb event shown to have a
massive binary companion to the progenitor. For SN~1993J, the
progenitor is thought to have been a 5.4~M$_{\odot}$ K~supergiant with
$\log L/{\rm L}_{\odot} = 4.5-5.5$, while the companion is now a
22~M$_{\odot}$ B2~Ia star of similar luminosity
\citep{nat04,svd02}. Both started out having initial mass
$\sim15$~M$_{\odot}$, but mass transfer to the companion caused their
evolutionary paths to diverge. In the case of SN~2001ig, no
pre-explosion imaging of sufficient quality exists to constrain the
progenitor spectral type, but as Paper~I discusses, the radio light
curve provides strong evidence for a WR progenitor, which typically
has $\log L/{\rm L}_{\odot} = 4.9-5.3$ \citep{lang91}. In close binary
systems involving a WR star and an OB companion, the mass of the WR
star can range anywhere from 5 to $>$50~M$_{\odot}$
\citep{tc03}, with much of this variation due to the extent of mass
transfer that has already occurred. The companion star to SN~2001ig
appears to have $\log L/{\rm L}_{\odot} \sim 4.5$, and
$M=10-18$~M$_{\odot}$ \citep{aq4} based on its SDSS colours and
inferred spectral type.

Although both events appear to have occurred within massive binary
systems, the optically-thin decline in radio flux was much smoother in
SN~1993J than in SN~2001ig. Hydrodynamical modelling by \citet{sp96}
indicated that the effects of varying CSM density on the radio flux
become more pronounced as the viewing angle approaches the orbital
plane. Although based on a small sample, this might explain why not
all Type IIb/Ib/Ic radio light curves exhibit such pronounced
structure, even if they all originated in WR + massive companion
binaries. \citet{03bg} found an uncanny resemblance between the radio
light curves of the Type Ibc SN~2003bg and those of SN~2001ig, even
down to having almost identical intervals between bumps.  They argue
that this is too much of a coincidence to be explained by the binary
viewing angle scenario, and propose instead that the progenitor
systems of both SNe are single WR stars undergoing enhanced mass loss
episodes every decade or so in the lead-up to the explosion. However,
they can offer no physical explanation for the cause of this
phenomenon or its periodicity, whereas the existence of a massive
binary companion as revealed here for SN~2001ig provides a natural and
simple means of modulating mass loss in the WR progenitor. Even where
pre-explosion archival data is not available, ongoing postmortem
investigations such as this one can still help us reconstruct the
circumstances surrounding the death of massive stars and whether
other parties were involved.


\section*{Acknowledgments}

Based in part on observations made with ESO Telescopes at the Paranal
Observatory; the Anglo-Australian Telescope; and at the Gemini
Observatory, which is operated by the Association of Universities for
Research in Astronomy, Inc., under a cooperative agreement with the
NSF on behalf of the Gemini partnership: the National Science
Foundation (United States), the Particle Physics and Astronomy
Research Council (United Kingdom), the National Research Council
(Canada), CONICYT (Chile), the Australian Research Council
(Australia), CNPq (Brazil) and CONICET (Argentina). This research has
made use of NASA's Astrophysics Data System Bibliographic Services
(ADS), as well as the NASA/IPAC Extragalactic Database (NED) which is
operated by the Jet Propulsion Laboratory, California Institute of
Technology, under contract with the National Aeronautics and Space
Administration.  We wish to thank S.~Allam and J.~A.~Smith for
providing SDSS magnitudes for T~Phe in advance of publication, as well
as T.~Matheson and A.~Filippenko for the spectra of SN~1993J.


\bsp

\label{lastpage}

\end{document}